\documentclass[aps,prl,twocolumn,groupedaddress,showpacs,amsmath,amssymb]{revtex4}
\usepackage{epsfig}
\usepackage{amssymb}
\usepackage{amsmath}
\usepackage{graphicx}
\usepackage{amsfonts}
\usepackage{epsfig}
\usepackage{revsymb}
\usepackage{bm}
\usepackage{latexsym}
\usepackage{hyperref}

\begin{document}

\title{Observation of the "burst-like growth" regime for the $^4$He crystals nucleated in a metastable liquid.}

\author{V. L. Tsymbalenko}
\email[]{vlt49@yandex.ru}
\affiliation{NRC Kurchatov Institute, 123182 sq.Kurchatov 1 \\
                  Institute for Physical Problems RAS, 119334 Kosigin st.2 \\ Moscow, Russia}


\begin{abstract}
The "burst-like growth" regime is observed for the $^4$He crystals with the growth defects. The observation has confirmed the hypothesis for the same  physical mechanisms responsible for the transition of the crystalline facets to the state of abnormally fast growth at high and low temperatures. The relaxation process of the kinetic growth coefficient is found to be similar to the relaxation of the elastic modules of the crystal at the end of the fast growth stage. The kinetic growth coefficients are determined at the stages of fast and slow growth. The crossover from the fast to slow kinetics of crystal facet growth is found to be drastic.
\end{abstract}


\maketitle

\section{Introduction}
In 1996 the amazing growth of helium crystals is discovered in two different series of experiments.  In the first series the phenomenon is observed in the quasi periodic growth of the perfect crystalline c-facet within the range 2-250~mK. This is the so-called "burst-like growth"  ~\cite{Finn01, Finn02}. Under constant liquid flow into the container the pressure increases since the  crystal facet does not grow due to lack of growth sources. However, as the supersaturation reaches the order of tenths of millibar, the crystal facet starts  to grow suddenly. The displacement of the facet is observed by the optical methods.  The fast growth results in a drastic pressure drop. Then the facet becomes immobile, the pressure enhances and the cycle repeats.

In the second series at temperatures 0.4-0.75K the $^4$He crystal nucleates in metastable superfluid liquid ~\cite{VLT01}. Within this temperature range the growth kinetics is governed by the most slowly growing sections of the surface, i.e., c- and a-facets.  The crystal grows as a hexagonal prism. After nucleation, the crystal grows slowly. The growth is accompanied with the slow pressure drop in the container. Then, the kinetic growth coefficient of all facets increases drastically by 2-3 orders of the magnitude.  As a result, the whole crystal grows rapidly within $\sim200\mu s$, Its growth  is accompanied by the drastic pressure drop. The shape and size of the crystal at this stage is recorded by video filming.

In the both series of experiments the phenomenon is observed when certain temperature-dependent magnitude of supersaturation $p_b$ is exceeded. The phase diagrams, showing the regions of normal and abnormal growth for the different temperature ranges, do not contradict each other. We observe the similar stochastic nature for the onset of the fast growth stage and for the effect of a small $^3$He impurity on the boundary supersaturation $p_b$. This similarity makes it possible to propose that the crystal growth acceleration under different experimental conditions leads to the same mechanism. If this hypothesis is correct,  the burst-like growth mode should be observed at the temperatures above 250~mK.
A review of experimental techniques, results obtained for phase diagrams $p_b(T)$ of the burst like growth, statistics of its nucleation, facets growth rates, elastic modulus relaxation of a crystal after the end of rapid growth, etc. is given in Ref.~\cite{VLT_UFN}.

To initiate the "burst-like growth" mode for the  equilibrium crystals at T=0.4-0.75~K, the supersaturation should be produced exceeding the $p_b$ boundary supersaturation equal to 2-10mbar at these temperatures.  An attempt to create such conditions in the dynamic manner has failed when the crystal is encircled with the superfluid helium, see Ref.~\cite{VLT_UFN}, Chapter~4.8. The next attempt is made for the crystals in the normal state.
 The point is that, in contrast to perfect crystals in experiments~\cite{Finn01, Finn02}, the facets of such crystals have growth defects~\cite{VLT1995}.
As a result, when liquid is pumped into the container, the crystal grows absorbing most of the inflowing liquid.  In addition, the high capillary impedance, the large volume of the container 4--10~cm$^3$, and the limited capabilities of the external pressure system do not allow  us to produce a high liquid flow sufficient to compensate for the crystal growth.  As a result, the  supersaturation required is not achieved.

This paper presents the results of measurements with a technique that   overcomes partially the limitations of the previous experiments. The internal volume of the container is reduced by a factor  $\sim$50 and the impedance of the inlet capillary is reduced by one order of magnitude.

\section{Experimental technique and the results}
The crystals are grown in a container according to the method described in Ref.~\cite{VLT_UFN}, Chapter 2. The internal volume of the container is 80~mm$^3$. A tungsten needle for the crystal nucleation in a metastable liquid $^4$He is placed at the center of the container. One of the container walls represents membrane of capacitive sensor with the time lag $160 \mu s$. The main difference between this technique and the previous one is that the nucleation of a crystal and its next growth take place under continuous liquid flow into the container. The experiments are carried out at two temperatures 0.49K and 0.74K. The upper limit of supersaturation is limited by spontaneous crystal nucleation on the inner wall of the container. In these experiments, the supersaturation magnitudes are within the range 0.1~--~5~mbar. The kinetic growth coefficient $K$ is determined by the expression
 \begin{equation}
V_{surf} = K\frac{\Delta\rho}{\rho\rho'}\Delta p
\end{equation}
where $V_{surf}$ is the surface growth rate, $\rho$ and $\rho'$are the densities of liquid and solid helium, $\Delta\rho = \rho' - \rho> 0$. The overpressure $\Delta p$ is measured from the phase equilibrium pressure.

The supersaturation $\Delta p_0$ at crystal nucleation determines its growth mode.
Figure 1 shows the magnitudes of the kinetic growth coefficient for the series of crystals at 0.74~K. It can be seen that the crystals can be separated into two groups.
\begin{figure}
\begin{center}
\includegraphics[%
  width=1.0\linewidth,
  keepaspectratio]{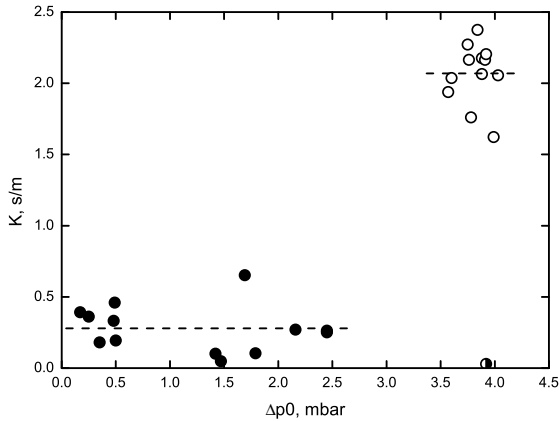}
\end{center}
\caption{Dependence of the average kinetic growth coefficient on the initial supersaturation $\Delta p_0$ at T = 0.74K. Open circles correspond to the crystals with anomalously fast growth. The solid circles are crystals with normal slow kinetics of the facet growth due to defects. The half-filled circle shows the magnitude of the kinetic growth rate calculated from the envelope $\Delta p(t)$. See Fig.2, upper graph, lower curve.}
\label{fig1}
\end{figure}
The crystals that start growing with the supersaturation less than $\sim$2.5~mbar
have the kinetic growth coefficient equal to $K = 0.28 \pm 0.16$s/m.
For the initial supersaturation larger than $\sim$3 mbar, the average kinetic growth coefficient is $2.07 \pm 0.20$s/m.  The boundary supersaturation $p_b$ lies within the range 2.5-3.5~mbar.
This magnitude is a half of the boundary supersaturation separating the regions of normal
and abnormal crystal growth studied earlier, see Fig.11 in Ref.~\cite{VLT_UFN}.
For T~=~0.49K, this boundary lies below 1~mbar which is also much
lower than the magnitude $p_b \approx 2$~mbar determined earlier.

\begin{figure}
\begin{center}
\includegraphics[%
  width=1.0\linewidth,
  keepaspectratio]{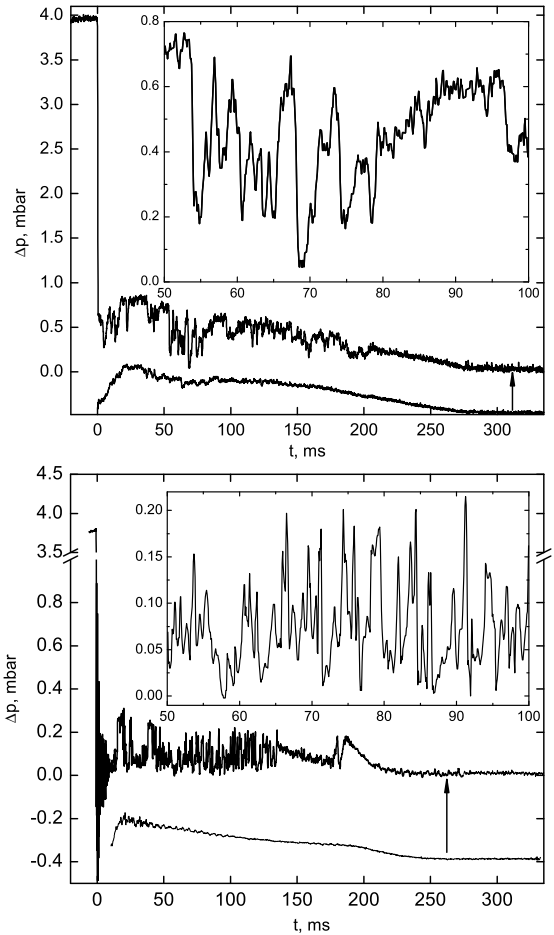}
\end{center}
\caption{Pressure changes in the container during crystal growth. The upper graph refers to a temperature of 0.74K, the lower one - 0.49K. The curves with jumps at t = 0 in both plots show the pressure change with the growth of a particular crystal. The smooth curve, shifted lower by the arrow, is the result of averaging over a series of measurements with the same starting conditions. The inset, on an enlarged scale, shows burst-like growth like pressure surges, Refs. ~\cite{Finn01, Finn02}.}
\label{fig2}
\end{figure}

Figure 2 shows the change in the pressure in the container during crystal growth at two temperatures.
At the moment t=0 the critical nucleus appears.
The crystal grows fast. The growth is accompanied by the drastic pressure drop  to the phase equilibrium pressure. For temperature 0.74K,
the supersaturation decreases gradually. At T = 0.49K the facet growth kinetics
enhances so much, entailing the oscillatory crystal growth which decays in $\sim$10~ms.
The evaluation for a ratio of the amplitude of the first oscillations period to the starting
pressure gives the magnitude $K$ = 6-7~s/m at this temperature.
The magnitudes of the kinetic growth coefficient and their temperature dependence
are consistent with the results obtained earlier~\cite{VLT_UFN}.

The liquid flow into the container forces the crystal to grow continuously.
If the kinetic growth coefficient of crystal facets is small, the growth occurs at significant supersaturation  $\Delta p$. Provided that the facets switch onto the "burst-like growth" mode with the kinetic growth coefficient $K$ larger by the orders of the magnitude, the supersaturation proves to be very small. The pressure records show both of these processes.
The return to the slow kinetics of crystal surface growth starts after completing
of the fast growth stage. The supersaturation increases, passes through a maximum in
the region of $\sim$30ms, and then vanishes at  $\sim$250ms.
This is clearly seen in the $\Delta p(t)$ curves obtained by averaging over a series of measurements.
The pressure records for the crystals nucleated under same conditions are aligned at the onset
of growth, summed up, and normalized. Such treatment averages the quasiperiodic pressure jumps.
The lower curves in Fig.2 correspond to the envelope of the $\Delta p(t)$ dependences.
Figure 3 shows the dependences of the kinetic growth coefficient $K$ calculated
by the method in Ref.~\cite{VLT2004}.
At first, the relaxation  kinetics is seen, corresponding to the rise in pressure at the averaged curves in Fig.2. The time constant at a temperature of 0.74K is 5~ms and at T = 0.49K is 7~ms.
After $\sim$30ms the kinetic growth coefficient becomes almost constant.
For temperature of 0.74K, its magnitude is $K = 0.030 \pm 0.006$~s/m, see the half-filled circle  in Fig.1. The cooling to 0.49K increases growth coefficient $K$ almost twice as large to  $K = 0.07 \pm 0.007$~s/m.

The records in Fig.2 show the pressure jumps in the interval 30-250~ms.
The sharp pressure drop evidences for the fast crystal growth.
The subsequent increase in supersaturation means a return to slow kinetics.
From the record $\Delta p(t)$ in Fig.2 one can see that the pressure jumps appear at the supersaturations lying within   range 0.3-0.8~mbar at temperature 0.74 K.
The decrease in temperature shifts this interval down to magnitudes 0.05-0.2~mbar.
The temperature affects the frequency of pressure jumps, as can be seen from Fig.2 in the insets.
 For T = 0.49K, the pressure jumps are formed more frequently than at T = 0.74K.
The shape of the pressure jumps is asymmetric.
After the onset of the burst-like growth state at T=0.74K,
the pressure drop occurs in $\sim 400 \mu s$.
This time is determined by the kinetic growth coefficient and corresponds
to the fast growth kinetics after nucleating the crystal.
The supersaturation returns to its original magnitude in 4-5~ms.
For temperature 0.49K, the pressure drop time is $200-250 \mu s$.
Note in this case that the time is not determined with the kinetic growth coefficient but with the frequency of soft modes of pressure oscillations during crystal growth ~\cite{VLT_UFN}, Chapter 2.5.
The supersaturation recovery time decreases to 0.6-0.8~ms. After $\sim$250~ms,
the supersaturation decreases and the burst-like growth process stops.
\begin{figure}
\begin{center}
\includegraphics[%
  width=1.0\linewidth,
  keepaspectratio]{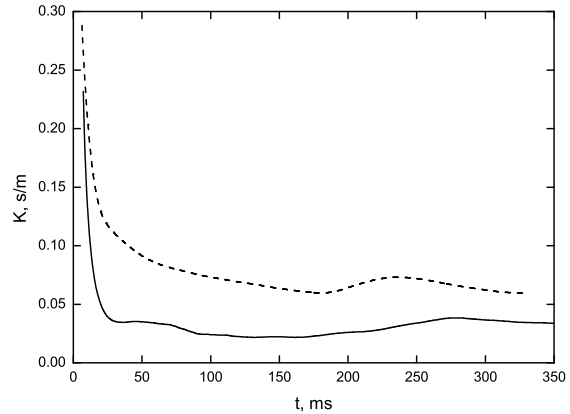}
\end{center}
\caption{Changes in the kinetic growth coefficient over time. The solid curve is calculated from the envelope of a series of records at T = 0.74K. Dashed line - along the series at T = 0.49K.}
\label{fig3}
\end{figure}

\section{Discussion}
The experiments have shown that the burst-like growth regime, observed previously for the perfect crystals, is also realized for the crystals with growth defects.
For high temperatures 0.4-0.75K and for low temperatures of 2-250~mK, the effect of anomalously fast crystal growth manifests itself in the same way.
This is a strong argument for the physical identity of the both effects.
Though this conclusion does not clarify the physical mechanism responsible for the effect, nevertheless it allows us to eliminate many possible explanations.
For example, since the effect is observed for the perfect crystal facet ~\cite{Finn01, Finn02},
all the possible explanations associated with the presence of growth defects should be disregarded.

The difference between the  present experimental setup and the previous one is the container volume.
The volume is reduced by $\sim 2$ orders of the magnitude.
The volume of the crystal is proportional to the internal volume of the container.
For this reason, the final crystal sizes in these experiments are $\sim$5~times as smaller than those of the crystals studied previously. It is possible that the shift of the boundary supersaturation $p_b$ is related to the crystal volume.

The magnitudes of the kinetic growth coefficient at the fast growth stage, calculated from the pressure drop after the crystal nucleation, agree well with the magnitudes measured before. That is, the fast growth of  the crystals, large and small in some meaning,  is similar.

The previous experiments performed within range 0.48-0.68K have only shown the  general features for returning the crystal facet growth kinetics to the normal regime ~\cite{VLT2004}.
It is shown that the significant relaxation takes place for time  $\sim$20~ms after ending the fast growth stage. Next, the slow relaxation is observed to the normal magnitudes of the kinetic growth coefficient. The details of this process have not been clarified due to experimental limitations indicated in the Introduction.

The reduction of the kinetic growth rate within $\sim$30~ms after the fast growth stage demonstrates the relaxation of the burst-like growth state to the normal state of low growth kinetics. Note that for the same time, the relaxation of the real and imaginary parts of the crystal elastic modulus takes place, see  Ref.~\cite{VLT_UFN}, Chapter 4.7.2. The relaxation time of this process changes insignificantly within the range 0.4-0.75K and takes 3-4~ms. Unfortunately, we do not know either this coincidence is occasional or this is two aspects of the same process since no answer can be obtained from the data available.

The magnitudes for the kinetic growth coefficient of the crystals grown in the normal state are one order of magnitude lower than the growth coefficients at the fast growth stage of fast growth, see Fig.1.  The growth coefficient measurements performed by a number of authors at about 0.75 K have a wide dispersion of these magnitudes from $4*10^{-4}$ to 0.02~s/m, see Fig.4 ~\cite{VLT_UFN}. The dispersion is not surprising since the crystal facets grow due to growth defects in this stage. The defect structure  depends essentially on the conditions of crystal growth, e.g. annealing, etc. The conditions are different in various experiments. The effect of different concentrations of growth defects on the kinetics is clearly seen in the different growth rates of the equivalent a-facets in the process of free crystal growth ~\cite{VLT1995}. It is possible that the high growth rates of "small" crystals are associated with their non-ideal structure. Note that the stationary magnitudes of the kinetic growth coefficient, calculated from the pressure envelope, are one order of the magnitude smaller than those for the normal crystal growth in these experiments, see Fig.3. The magnitudes are close to those measured previously in the works of other authors

To summarize, we have succeeded  to reproduce the burst-like growth mode in the crystals containing the growth defects. This means that the physical mechanisms responsible for the crossover of the facets to the state of abnormally fast growth are identical. The relaxation process for the kinetic growth coefficient is found to be similar to the relaxation of the crystal elastic modules at the end of the fast growth stage. The kinetic growth factors at the  fast and slow growth stages have been determined. The crossover from the fast to slow kinetics is found to be drastic as well.

\section{ACKNOWLEDGMENTS}
The author is grateful to V.~V.~Dmitriev for the possibility of performing these experiments at Kapitza Institute for Physical Problems RAS. The author is also grateful to V.~V.~Zavyalov for supporting this work, S.N.Burmistrov for helpful comments and V.~S.~Kruglov for interest to the work.


\begin{thebibliography}{99}

\bibitem {Finn01} { A.~V.~Babkin, P.~J.~Hakonen, A.~Ya.~Parshin, J.~S.~Penttila, J.~P.~Ruutu, J.~P.~Saramaki, G.~Tvalashvili}, {\it Phys.Rev.Lett.}, {\bf 76}, 4187 (1996)
\bibitem {Finn02} { A.~V.~Babkin, P.~J.~Hakonen, A.~Ya.~Parshin, J.~P.~Ruutu, G.~Tvalashvili},  {\it J.Low Temp.Phys.}, {\bf 112}, 117 (1998)
\bibitem {VLT01} { V.~L.~Tsymbalenko}, {\it Phys.Lett.A}, {\bf 211}, 177 (1996)
\bibitem {VLT_UFN} { V.~L.~Tsymbalenko}, {\it Physics-Uspekhi}, {\bf 58}, 1059 (2015)
\bibitem {VLT1995} { V.~L.~Tsymbalenko}, {\it Ukr.Low Temp.Phys.}, {\bf 21}, 120 (1995)
\bibitem {VLT2004} { V.~L.~Tsymbalenko}, {\it JETP}, {\bf 99}, 1214 (2004)

\end{thebibliography}
\end{document}